\documentclass[12pt]{article}

\usepackage{natbib,url}
\usepackage{enumitem}
\usepackage{subcaption}

 \newcommand{\ket}[1]{|#1\rangle}
 
 \newcommand{\bkt}[2]{\langle#1|#2\rangle}

 \newcommand{\mc}{\mathcal}
 \newcommand{\bibsuffix}[1]{}

 \title{On the Status of Quantum State Realism}
 \author{Wayne C. Myrvold \\
 Department of Philosophy \\
 The University of Western Ontario \\
 wmyrvold@uwo.ca}
 \date{Forthcoming in \emph{Scientific Realism and the Quantum}, \\ Steven French and Juha Saatsi, eds., \\ Oxford University Press, 2020.}

\begin{document}

 \maketitle
 \abstract{In this chapter,  I argue that we have good reason for being realist about quantum states.  Though a research programme of attempting to construct a plausible theory that accounts for quantum phenomena without ontic quantum states is well-motivated, that research programme is confronted by considerable obstacles.  Two theorems are considered that place restrictions on a theory of that sort: a theorem due to Barrett, Cavalcanti, Lal, and Maroney, and an extension, by the author, of the Pusey Barrett Rudolph theorem, that employs an assumption weaker than their Cartesian Product Assumption. These theorems have assumptions, of course. If there were powerful evidence against the conclusion that quantum states correspond to something in physical reality, it might be reasonable to reject these assumptions. But the situation we find ourselves in is the opposite: there is no evidence at all supporting irrealism about quantum states.}

\section{Introduction}  There is a long tradition, very much alive in the present day, of irrealism  about quantum states---that is, of denying that quantum states represent anything in physical reality.\footnote{This tradition goes back at least as far as Bohr and Einstein, who  agreed that quantum mechanics should not be taken as descriptive of physical reality, though they disagreed on the propriety of  seeking a theory that would be.  For Bohr, all description of physical reality had to be couched in classical terms, and the limits of classical physics were the limits of physical description.  Einstein argued, in several places (see, \emph{e.2g.}, \citealt{EinsteinPR}), that quantum states should be regarded as akin to the probability distributions of classical statistical mechanics, that is, as representing incomplete knowledge of some deeper underlying physical state.    Contemporary representatives of this tradition include those who call themselves \emph{QBists} (formerly ``Quantum Bayesians'') \citep{CFSQB,FMSQB,FSQB}.  The central tenet of QBism is that a quantum state assignment is nothing more than a way of encoding an agent's  subjective degrees of belief about that agent's own future experiences.  Views that take quantum states to be representations of a state of knowledge, rather than physical reality, are often called \emph{$\psi$-epistemic} views.  A prominent exponent of an epistemic view of quantum states is Rob Spekkens (\citeyear{SpekkensToy,SpekkensPsiOnt}). Richard  Healey(\citeyear{HealeyPragQM,HealeyBridges,HealeyRevolution,HealeyPQR}) advocates a pragmatic view of quantum states, which denies that quantum states are representational.  See \citet{sep-quantum-bayesian} for an overview of views of this sort.}

In this chapter, I will argue that the grounds we have for taking quantum states to represent physical properties of the systems to which they are ascribed are  as strong as the grounds we have for taking atoms or electromagnetic waves to be real and to have something like the properties we ascribe to them. I will take it for granted that we do, indeed, have sufficient grounds for belief in the reality of atoms and electromagnetic waves. It is not my intention to try convince a committed scientific anti-realist to make an exception for quantum states. The issue at hand is orthogonal to the age-old struggle between scientific realists and anti-realists.  My targets here are  those who deny that quantum states represent anything in physical reality  from a standpoint that holds  that one can, indeed, under certain circumstances ascribe reality to entities that are not directly observable, but  take it that there are  reasons specific to the quantum context for denying ontological import to quantum states.

The question of the representational status of quantum states  is a question that can be addressed even though we know  that quantum mechanics is not a fundamental theory, but rather, a non-relativistic, low-energy approximation to a quantum field theory, and even though we have good reason to believe that even our best quantum field theories are effective theories, low-energy approximations to some deeper theory that would incorporate gravitation.  Electromagnetic fields are real, even though classical electromagnetic theory is an approximation, valid within a limited regime, to a more fundamental theory.  Classical physicists had good reason to believe that any deeper theory would include electromagnetic fields in its ontology, even if these fields are not precisely as classical electromagnetic theory conceives them to be. A successful argument that quantum states are real would not be one that depended crucially on  a fiction that quantum mechanics is exactly right.   What is required is an argument that we can expect any theory that recovers the predictions of quantum mechanics, or at least a close approximation to them, within the known domain of applicability of quantum mechanics, to have something corresponding to quantum states in its ontology, either as fundamental ontology or emergent from something more fundamental. As we shall see, this imposes a non-trivial constraint, as it would not do to take as a premise of the argument some condition that is violated by quantum field theory.

In section \ref{PsiOnt} we will examine some theorems that circumscribe the realm of possible physical theories that can account for quantum phenomena.  The first to be considered is the result of Barrett, Cavalcanti, Lal, and Maroney (\citeyear{BCLM}), which shows that quantum states cannot be construed as some have hoped they could be, as probability distributions over an underlying state space, in such a way that operational indistinguishability of quantum states can be accounted for in terms of overlap of the corresponding probability distributions.  We will then consider the \emph{  }theorem of Pusey, Barrett and Rudolph (\citeyear{PBR}), which demonstrates, on the basis of an assumption about independent preparations performed on distinct systems, known as the Preparation Independence Postulate (PIP), that distinct pure quantum states are ontologically distinct.\footnote{In labelling the Preparation Independence Postulate and its relatives, we follow the terminology of \citet{LeiferPsiOnt}.}

In accordance with the requirement that the ontological lessons we draw from physical theory  rely only on premises that can reasonably be expected to be preserved under the transition to a successor theory, we should ask  whether the PIP passes muster in that respect.  And, indeed, there is an aspect of it that is problematic, in light of quantum field theory. The Postulate assumes that, for a system consisting of two or more spatially separated subsystems,  for appropriate preparations the resulting state of the whole can be regarded as consisting merely of a list of states of the component subsystems.  This assumption is called the Cartesian Product Assumption (CPA).  This holds in classical physics, but is violated  in any theory that is realist about quantum states.  It holds within a fragment of quantum mechanics in which the states prepared are product states (that is, in which there is no entanglement between the spatially separated parts).  However, quantum field theory gives us  incentive to doubt whether product states can reliably be prepared (see \citealt{EntangOpenSystems}).

This gives us motivation to formulate a substitute for the PIP that does not invoke or presuppose the CPA.  In  section \ref{thm}, I present a condition that holds whenever the PIP holds but which is strictly weaker than it, which I call the  \emph{Preparation Uninformativeness Condition} (PUC).  This condition requires no assumption about the structure of the state space of composite systems or its relation to subsystem state spaces.  On the basis of the assumption of this condition, it can be shown that distinct quantum states---as long as they aren't too close together---must be ontologically distinct.

All of these theorems are couched within the ontological models framework.  This framework, explicitly formulated by \citet{SpekkensHarrigan}, codifies reasoning implicit in the practice of information theory, quantum or otherwise, and, indeed, in much of science and everyday life.  Aspects of the framework could be rejected, but, as the sort of reasoning invoked is implicit in so much of science, strong grounds would be required for doing so.   Now, it is, of course, possible that the methods of inference that we routinely employ in other domains of science lead us astray when it comes to investigating the quantum domain. One thing that is, after all, uncontroversially true is that any realist construal of quantum mechanics  entails rejection of some one or other tenet of classical physics that one might have otherwise thought could be taken for granted.  I acknowledge this, and, indeed, I accept that, if we had strong evidence that these methods of inference lead us astray when applied to the quantum domain, it would be reasonable to reject them.  What is \emph{not} reasonable, and not consistent with an earnest investigation of the world around us, is to reject methods of inference simply because their application would lead to conclusions that one finds unpalatable.  The claim I am advancing in this chapter is that we do not have grounds for doubting the conclusion that quantum states represent something in physical reality that are sufficient to undermine the premises and modes of reasoning that lead to that conclusion.

\section{Arguments for anti-realism about quantum states} First, let us look at  some of the reasons that have been given for denying that quantum states represent something physically real.     There are two ways that one could take these.  One could  take them  as motivating pursuit of a project of trying to develop a theory in whose ontology quantum states do not appear.  Another way would be to take them as arguments for the conclusion that quantum states do not represent anything physically real.  The difference matters, because the criteria for success of the arguments are different, depending on what the conclusion is taken to be.    Upon undertaking a research project, say, to attempt to find a theory of a certain sort, one does not require assurance that the project can reach its goal.  All that is needed is that it appears to be a promising line of research, whose goal, if reached, would constitute an advancement in understanding.  Moreover, unsuccessful attempts themselves may lead to deeper understanding, if they help us to understand why they were unsuccessful---especially if we learn that the goal could not be reached.

If, on the other hand, we had strong evidence for the conclusion that  quantum states are not representational, then, faced with arguments, such as those to be considered later, that they are, this evidence might afford us reason to be suspicious of, and perhaps even reject, the premises that lead to the conclusion.  I do not think we are in such  a position.  The sorts of arguments that are given for anti-realism about quantum states serve their purpose well if they are taken to provide motivation for a certain line of theory pursuit.  If, however, one were to take them as providing evidence for the conclusion that quantum states do not represent anything physically real, the evidence provided is  weak at best, and certainly not sufficient to cast doubt on mundane assumptions that otherwise would be accepted without question.

In my opinion, the project of constructing an empirically adequate physical theory whose ontology would dispense with quantum states was, indeed, a worthwhile and well-motivated project, and, moreover, one that has been fruitful, precisely because it has led to deeper insight into the obstacles that such an endeavour faces. The situation bears some resemblance to the question of the viability of local hidden-variables theories. Inspired by the EPR argument for the incompleteness of quantum mechanics, J. S. \citet{BellHV}  raised the question of whether there could be a hidden-variables theory that did not share the nonlocality of the de Broglie-Bohm theory, noting that, as far as he knew, there was ``no \emph{proof} that \emph{any} hidden variable account of quantum mechanics \emph{must} have this extraordinary character.'' What happened was that the quest for a local hidden-variables theory led to an impossibility proof.  Bell's proof rests on assumptions, as any proof must. One of these is the so-called ``no-conspiracy'' assumption, namely, that it is possible to create an experimental set-up in which the instrumental settings are effectively independent of the prepared state of the system to be experimented upon.  One can, without logical contradiction, reject this assumption.  But a rejection of this sort is a blunt instrument; it could be used to reject \emph{any} experimental conclusion one doesn't like.  The relevant question is: in the case of the Bell experiments, do we have evidence of conspiracies of this sort?  If the answer is simply that they are being invoked to avoid an unwelcome conclusion, then it seems not unfair to say that those who invoke them have abandoned the sincere quest for knowledge about the world.

\citet{LeiferPsiOnt} has helpfully compiled some of the chief arguments that have been advanced in favour of rejecting realism about quantum states. Leifer regards these as sufficiently strong that a reader who appreciates their force should find the $\psi$-ontology theorems surprising.\footnote{``$\psi$-ontology,'' with pun intended, is a term that has gained currency among physicists who discuss these matters for views that hold that quantum states represent something in physical reality. It is attributed to Christopher Granade, who was a student  in Rob Spekkens' quantum foundations course at Perimeter Institute for Theoretical Physics in 2010 (see \citealt[p. 71]{LeiferPsiOnt}). }  As already  mentioned, I think that these considerations are better thought of as providing motivation for a project of constructing  a theory that does not include quantum states in its ontology, rather than as positive evidence for the unreality of quantum states.

The first argument Leifer considers stems from the Rob Spekkens' toy theory \citep{SpekkensToy}.  With a remarkably simple construction, Spekkens demonstrates that a number of phenomena that we might have thought of as distinctively quantum can be captured by a model that is essentially classical, with restrictions on state preparation and on access to information about the state of the system. An elementary system in this toy theory is a set of four boxes with a ball that can be in one of them.  The preparable states of individual systems are restricted in such a way that the most one could know is that the ball is in one of two of the boxes, with equal probability for each.  For a pair of elementary systems, in addition to the product states, there are also entangled states, in which, for each of the subsystems, the ball is equally likely to be in each of the boxes, but there is perfect correlation between the two systems.

Features of quantum theory that are reproduced in the toy theory include existence of pure states that cannot reliably be distinguished, no-cloning, and (an analogue of) interference. Quantum phenomena  that are provably not reproduced in the toy theory include violations of Bell inequalities, and the possibility of obtaining a Kochen-Specker obstruction.\footnote{If this term is unfamilier to you, see \citet{sep-kochen-specker} for background.} It is suggested that the quantum-like phenomena that the toy theory reproduces are evidence that quantum theory itself is a theory of the type instantiated by the toy theory, that is, a theory with an essentially classical state space and a restriction on possible state preparations.  An alternative moral that could be drawn is that  it was a mistake to think of those phenomena as distinctively quantum \citep[p. 182]{MyrvoldBoston}; such phenomena are, at most,  \emph{weakly} nonclassical  \citep[p. 92]{SpekkensQQ}.

Support for this latter moral is found in  work on generalized probabilistic theories.  The  framework of generalized probabilistic theories encompasses a wide variety of different sorts of probabilistic theories.\footnote{See \citet{JH2014} for an introduction and overview.} The scope of this framework is wide enough to include classical probabilistic theories, with or without restrictions on state preparations, quantum theories, and, in addition, a whole host of theories neither classical nor quantum.  The theory whose state space consists of the states allowed in Spekkens toy theory and probabilistic mixtures of these states falls within the scope of this framework.

Within the class of generalized probabilistic theories, there is a distinguished class, those with a  state space that  is a  simplex, meaning: any mixture has a unique decomposition into pure states. These theories are the classical theories with no restrictions on state preparation.  Call these \emph{fully classical} theories, to distinguish them from classical theories with preparation restrictions, which exhibit some features usually thought of as non-classical.   Theories that are not fully classical  have in common the feature of having pure states that are not distinguishable.\footnote{Provided that  the theory's state space is the convex hull of its set of pure states.}  \emph{Ipso facto} they have all of the consequences of that condition, such as no-cloning.  It is no surprise that quantum theory, being one of the theories that are not fully classical, shares with the Spekkens toy theory the features that are shared by all theories that are not fully classical.  This is no reason to think that there should be some commonality in the physical interpretation of all such theories.

Leifer suggests that the fact that quantum theory falls within this broad framework, which also includes classical probability models, is evidence for an epistemic view of quantum states.  ``In this theory, quantum states are playing the same role in the quantum case that probability measures play in the classical case, and so it is natural to interpret quantum states and classical probabilities as the same kind of entity'' \citep[p. 76]{LeiferPsiOnt}.  Thinking along these lines, it  seems to me, fails to do justice to the generalized probabilistic theories framework. The framework was constructed to embrace a wide range of theories, and arguably what it presumes is just the minimum one would expect of any physical theory.  It is \emph{completely} neutral as to the physical interpretation of the states of the theory, and in this neutrality is its strength, as it is the source of  generality.

The second argument considered by Leifer concerns fragments of quantum theory that can be recovered in a classical model.  For instance, under suitable restrictions on quantum state preparations and evolutions, the Wigner function, a function on classical state space definable in terms of the quantum state, is positive and acts like a density function for a probability distribution over a classical phase space.

Is this surprising, on the assumption that quantum states are real?  I think not.  Suppose that you had never heard of quantum theory, but had become convinced that classical physics is inadequate in certain ways, because its predictions in certain domains are incorrect.  You would, quite reasonably, expect any successor theory to recover the successes of classical physics. This means that you would expect to obtain something like classical behaviour in the relevant domains, or, to put it another way, that there be fragments of the theory that exhibit classical or quasi-classical behaviour. Studies of the classical limit of quantum mechanics take positivity of the Wigner function as an indication of classicality. That quantum mechanics exhibits classical-like behaviour in certain domains---that is, that quantum theory has a quasi-classical limit---is not evidence that quantum states are unreal, but, rather, a precondition for taking quantum mechanics as a serious candidate for a comprehensive theory.

Leifer also takes, as a strength of the epistemic view of quantum states, the fact that it bypasses the notorious quantum state collapse.  Certainly, it is an attractive idea, one that has no doubt occurred to many, that collapse of the quantum state be thought of as nothing more than updating of information upon learning the result of a measurement.  The question is whether this can be made to work. Any approach to the so-called ``measurement problem,'' including one that denies that quantum states represent physical reality, owes us an account of what happens in an experiment.  The mainstream approaches---hidden-variables theories, dynamical collapse theories, and Everettian interpretations---all provide such accounts. Each of these deals with collapse in different ways. On the de Broglie-Bohm pilot wave theory,  collapse of the effective wave function is a demonstrable consequence of the theory.  On dynamical collapse theories, collapse is a real physical process.  And Everettian theories can explain why, under appropriate circumstances, agents may be justified in disregarding other branches of the wave function other than their own, just as if there had been collapse.   There is yet no worked-out proposal for a theory that embraces quantum phenomena on which quantum states are epistemic. At best we have  is a hope that an account of what happens during an experiment could be given on which quantum states play no part in the ontology.  The situation, then, is that all  of the main avenues of approach have rejected the problematic textbook version of collapse, with its reliance on a distinction between ``measurements'' and other physical processes, and have provided a unified account of the goings-on of physical systems that makes no use of this distinction at the fundamental level and which nevertheless gives an account of why the textbook formulation works as a heuristic. In each case, this is accomplished by taking the quantum state as part of the ontology.   Against this, the $\psi$-epistemicist offers only a hope that it could be accomplished without ontic quantum states.  As long as this hope remains unfulfilled, consideration of the issues surrounding the measurement problem remain a problem to be solved by, rather than evidence in favour of, a $\psi$-epistemic view.

To sum up: if there were powerful evidence against the conclusion that quantum states correspond to something in physical reality, it might be reasonable to question the assumptions behind the arguments, to be considered in the next section, for the reality of quantum states.  But the situation we find ourselves in  seems to be the opposite; there is \emph{no evidence at all} supporting  irrealism about quantum states.  At best we have considerations that suggest the pursuit-worthiness of the project of  attempting to construct a plausible theory that accounts for quantum phenomena without ontic quantum states.

\section{Arguments for an ontic construal  of quantum states}\label{PsiOnt}
The arguments for $\psi$-ontology will be framed against the background of the ontological models framework.  We will introduce this framework, then consider some arguments for  $\psi$-ontology.  The conclusions will differ in strength, depending on the strength of the auxiliary assumptions involved.   We will first consider the theorem of Barrett, Cavalcanti, Lal, and Maroney (\citeyear{BCLM}), which shows that indistinguishability of quantum states cannot be fully accounted for by overlap of probability distributions over an ontic state space. We will then consider the theorem of Pusey, Barrett, and Rudolph (\citeyear{PBR}), and then a variant of it that replaces the key assumption of this theorem, the Preparation Independence Postulate, with a weaker assumption, which we will call the Preparation Uninformativeness Condition (PUC).

\subsection{The ontological models framework}   Consider the following set-up, the sort of scenario with which information theory, be it quantum or classical, routinely deals.  Alice has a message that she wants to convey to Bob.  She has a physical system that she can send to him, after subjecting it to one of some set of available preparation procedures.  Alice and Bob have an agreed-upon coding that associates possible messages with the available preparation procedures.  Alice chooses her preparation procedure, subjects the system to it, and sends it to Bob, who performs an experiment, and takes the outcome as informative about what Alice has done.

In the simplest case, suppose that there are two procedures, which we will call $P_1$ and $P_0$,  and that Bob has available to him an experiment  with two possible results, $R_1$ and $R_0$, such that with probability one he obtains $R_1$ if Alice has performed $P_1$ and $R_0$ if she has performed $P_0$.  In such a circumstance, it is hard to escape the conclusion that Alice's preparation has an effect on the state of the mediating system, and that the sets of   states that could result from $P_1$ and $P_0$, respectively, are disjoint.  Insisting that all probabilities be regarded as subjective judgments, as QBists do,\footnote{See, in particular, \citet{CFS2007}.} does not change the situation.  If Bob's credences, or subjective degrees of belief, are such that he assigns probability one to the outcome $R_1$ conditional on the supposition of $P_1$, and  probability one to the outcome $R_0$ conditional on the supposition of $P_0$, the natural explanation for this is that Bob believes that Alice's preparation has an effect on the state of the system that influences the outcome, and that he believes that the sets of states that could result from $P_1$ and $P_0$, respectively, are disjoint.

Could this conclusion be avoided?    One could postulate that Alice's preparation has a direct influence on the result of Bob's future experiment, an influence unmediated by any influence on the state of the world between the preparation and the experiment.   One could also stipulate that we are forbidden to theorize about the states of the mediating system.  Moves of this sort would undermine the usual patterns of reasoning that underlie information theory, which presume that Bob, by doing an experiment on the system transmitted to him by Alice, gains information about what Alice did, mediated by the system that passed between them. One, could, perhaps imagine situations in which we had strong evidence for the unreliability of such patterns of reasoning, evidence strong enough to warrant rejecting them. \emph{Perhaps}!  But it should be noncontroversial that the mere fact that application of such inference schemes leads to the conclusion that the world is fundamentally nonclassical, or that it has features that some find unpalatable, does not constitute evidence for their unreliability.

 It doesn't change matters much if we stipulate, as QBists do \citep{FMSQB}, that Alice is forbidden to even consider the effects of her choices of on the probabilities of outcomes of an experiment performed by another agent.  To make a stipulation of this sort is to abandon the very framework of information theory, but it doesn't block the inference, as Alice can send messages to her future self, as an aid to memory.  Unless Alice believes that, when she looks tomorrow at the laptop  she typed  on  today, she will gain information about what she wrote earlier, mediated by the effect  on the internal state of the laptop of her choices made today, then it is hard to understand what she is doing, or why.

In cases in which the probabilities are different from zero and one, the reasoning is similar. Suppose that Alice has a choice between two coin-flipping procedures: $P_0$, which yields \emph{heads} and \emph{tails} with equal probability, and $P_1$, which yields \emph{heads} with probability $2/3$ and \emph{tails}  with probability $1/3$.  Alice chooses a preparation, flips the coin, and then passes it to Bob, who looks at it and sees  \emph{heads} or \emph{tails}. He thereby gains information about the preparation procedure. If his prior credences in $P_0$ and $P_1$ are $Cr(P_0)$ and $Cr(P_1)$, respectively, and if his conditional credences $Cr(H|P_0)$, $Cr(H|P_1)$, are those just mentioned, then, in the event of seeing \emph{heads},  an application of Bayes' theorem yields the result that his posterior credences in the two preparations should satisfy,
\begin{equation}
\frac{Cr(P_1|H)}{Cr(P_0|H)} = \frac{Cr(H|P_1) \: Cr(P_1)}{Cr(H|P_0) \: Cr(P_0)} = \frac{4}{3} \left(\frac{Cr(P_1)}{Cr(P_0)}\right).
\end{equation}
That is, his credence in $P_1$ is increased, and his credence in $P_0$ diminished, in such a way that their ratio is increased by a factor of $4/3$.  In the event of seeing \emph{tails}, his posterior credences in the two preparations should satisfy,
\begin{equation}
\frac{Cr(P_1|T)}{Cr(P_0|T)} = \frac{Cr(T|P_1) \: Cr(P_1)}{Cr(T|P_0) \: Cr(P_0)} = \frac{2}{3} \left(\frac{Cr(P_1)}{Cr(P_0)}\right).
\end{equation}
Thus, the result of looking at the coin is, for Bob,  informative about the preparation Alice chose; seeing \emph{heads} boosts his credence in $P_1$, and lowers his credence in $P_0$, whereas seeing \emph{tails} boosts his credence in $P_0$, and lowers his credence in $P_1$.

In this case, there are two disjoint classes of physical states that the coin can be in, corresponding to \emph{heads} and \emph{tails}.  Corresponding to each preparation are probabilities for the state of the coin ending up in each of these classes, and it is assumed that Bob can ascertain which of these classes the state of the coin is in without error, simply by looking.  This latter assumption is inessential. We might assume that Bob has some  blurriness of vision, which introduces error at the readout stage, and that, in any state of the coin, there is some probability that he will see it as \emph{heads}, and some probability that he will see it as \emph{tails}.   This changes little; as long as the net probability that he will see the coin as \emph{heads} is higher given preparation $P_1$ than it is given preparation  $P_0$, seeing heads should boost his credence that the preparation was $P_1$.

In reasoning of this sort there are two sorts of probabilities to be taken into account.  We have \emph{preparation probabilities}: we associate with each preparation procedure  a probability distribution over the possible physical states that could result from the preparation. We also have \emph{outcome probabilities}: for any experiment that can be performed on the system, for any physical state of the system we have, for each outcome, a  probability of obtaining that outcome given that physical state.

It is worth noting that \emph{nothing at all} in this sort of reasoning depends on whether these preparation and outcome probabilities are taken to be epistemic, or a matter of physics, or some mixture of epistemic and physical considerations. Bob might take it that the underlying physics is deterministic and that any uncertainty he might have about the result of a preparation stems from uncertainty about the details of what goes on in the preparation; all that matters  is that the choice of preparation matters to his credences about the resulting state in a way that matters to his credences about the results of his experiment.  If, on the other hand, the preparation and outcome probabilities are regarded as objective chances, then, provided that Bob knows what objective chance distribution to associate with a given preparation, and what distribution to associate with outcomes of a given experiment, and provided that his credences satisfy the Principle Principal,\footnote{This is a principal often tacitly assumed in probabilistic reasoning, which was explicitly identified and named by David \citet{LewisSGOC}.  It requires a meshing between an agent's degrees of belief in a proposition $A$ and her degrees of belief in propositions about possible chances of $A$.  The Principle requires that an agent's degree of belief in $A$, conditional on the supposition that the chance of $A$ is equal to $x$, be itself equal to $x$. } his conditional credences will be such that he takes the outcomes of the experiment to be informative about the preparation.  Those who wish to construe the probabilities that appear in the theorems to be considered below as purely epistemic are welcome to do so; the conclusion they will arrive at is that an agent whose credences about experimental outcomes conform to quantum mechanics ought to regard distinct quantum states as ontologically distinct.

These sorts of considerations, which have been more or less implicit in much of the discussions concerning the reality of quantum states, have been explicitly formulated by \citet{SpekkensHarrigan}.  We associate with any physical system a  physical state space, or \emph{ontic state space} $\Lambda$, and a set $\mc{L}$ of subsets of $\Lambda$ that will be taken to be the measurable sets, that is, the ones that are candidates for ascribing a probability to.\footnote{We make the assumption, which is usual in probability theory, that $\mc{L}$ contains $\Lambda$  and is closed under complementation and countable union; that is, we will take  $\mc{L}$ to be a $\sigma$-algebra.  See standard texts, such as \citet{Billing}. } With any preparation procedure $\psi$ is associated a probability distribution $P_\psi$ on the measurable space $\langle \Lambda, \mc{L} \rangle$.  For any  experiment $E$, with a set of outcomes $o_k$, $k = 1, \ldots, n$, there is a corresponding set of response functions $f_k$, such that $f_k(\lambda)$ is the probability of obtaining outcome $o_k$ in ontic state $\lambda$.  As these are probabilities for a set of mutually exclusive and jointly exhaustive alternatives, they must add up to one.  Thus, for all $\lambda \in \Lambda$,
\begin{equation}
\sum_{k = 1}^n f_k(\lambda) = 1.
\end{equation}
With these in place, the probability that a system subjected to preparation $\psi$ yields outcome $o_k$ when experiment $E$ is performed on it is the expectation value of $f_k$ with respect to $P_\psi$.
\begin{equation}
P_\psi(o_k) = \int_\Lambda f_k(\lambda) \, dP_\psi(\lambda)
\end{equation}

A few words of comment about the notion of preparation being invoked here.  Note that we associate with each preparation procedure  a corresponding probability distribution on the ontic state space. These probability distributions differ for different preparations, but it is assumed that, once the preparation performed has been specified, everything relevant to probabilities of outcomes of subsequent experiments has been specified.  In particular, a preparation screens off such things as details of the past of the system that are not relevant to specification of the preparation.  One could take this to be part of the meaning of ``preparation''---if you think that the past of a system continues to be relevant to future events, after a certain procedure has been performed, then you should take differing pasts to correspond to different preparations.  Local preparations---that is, preparations taking place in a bounded region of space and time---are taken to screen off correlations between the prepared systems and the world outside.

The assumption that preparations of this sort are possible, and, indeed, are routinely performed in laboratories, is a substantive assumption, an assumption that does not follow from anything like a condition of causal locality. It is neither necessarily true nor knowable \emph{a priori}.  However, it \emph{is} an assumption that lies deep at the heart of virtually all experimental science.  If we ever come to a point at which we have reasons to doubt this sort of assumption, it will not come about as a result of experiments that presuppose it.  And if we were presented with a reason to doubt this sort of assumption, it is hard to see how this doubt could be sufficiently contained so as not to undermine all of experimental science.  Fortunately, we are not in such a position, as no-one has offered grounds for doubting it.

Two preparations $\psi, \phi$ are said to be \emph{distinguishable} if and only if there is an experiment $E$ such that, for each  outcome $o_k$ of the experiment, either $P_\psi(o_k)$ is zero or  $P_\phi(o_k)$ is zero.  This means that, if a system is subjected to one of two preparations, but you don't know which, you can become certain of which it was by performing the experiment, as every outcome precludes one or the other of the possible preparations. This generalizes to larger sets of preparations: a finite  set of preparations $\{\psi_i \}$ is said to be {distinguishable} if and only if there is an experiment $E$ such that, for each outcome $o_k$, at most one of $\{P_{\psi_i}(o_k)\}$ is nonzero. Following \citet{LeiferPsiOnt}, we will say that a finite set of preparations  $\{\psi_i \}$ is \emph{antidistinguishable} if and only if there is an experiment $E$ such that each outcome of $E$ has zero probability on some preparation in the set.  That is, no matter what the outcome of the experiment $E$ is, it rules out at least one of the preparations.

Two preparations are said to be \emph{ontologically distinct} if there is a measurable subset $S$ of the ontic state space such that $P_\psi(S) = 1$ and $P_\phi(S) = 0$.  It is a straightforward theorem that any distinguishable set of preparations is pairwise ontologically distinct. The converse might not hold; a pair of ontologically distinct preparations might not be distinguishable. This will be the case whenever there are limitations on what one can learn about the ontic state of a system in a single experiment.

If a set of preparations $\{\psi_i \}$ is antidistinguishable, this entails that the corresponding probability distributions $\{P_{\psi_i}\}$ have null joint overlap.  That is, there is no subset $S$ of the ontic state space such that $P_{\psi_i}(S) > 0$ for all $\psi_i$ in the set.

In the coin-flip example, the two preparations, involving differing nonextremal probabilities for the outcomes \emph{heads} and \emph{tails}, are neither distinguishable nor ontologically distinct.  In the absence of limitations on available experiments, on a classical theory, any pair of ontologically distinct states will be distinguishable. What I mean by this is: if, for every measurable subset $S$ of the state space, there is an experiment that determines whether or not the state is in $S$, then every pair of ontologically distinct states is distinguishable.

In quantum mechanics, as is well-known, nonorthogonal states are not distinguishable.  If a pure quantum state is part of the ontology, then preparations of distinct pure states will be ontologically distinct, and so there will be ontologically distinct preparations that are not distinguishable.

The question arises whether nonorthogonal quantum states are analogous to classical states, in which indistinguishability of preparations corresponds to overlap in the associated probability distribution on the ontic state space.  If this is the case, then one and the same ontic state is compatible with distinct quantum states, which is to say: the ontic state does not uniquely determine the quantum state.  If, on the other hand,  the quantum state supervenes on the ontic state, then preparations corresponding to distinct pure quantum states will be ontologically distinct.  If an ontological model of quantum state preparations and experiments  is such that, for any two distinct quantum states $\ket{\psi}$, $\ket{\phi}$,  any pair of preparations $\psi$, $\phi$ that prepare those states are ontologically distinct, the model is said to be \emph{$\psi$-ontic}.  Harrigan and Spekkens  (\citeyear{SpekkensHarrigan}) define \emph{$\psi$-epistemic} as the negation of $\psi$-ontic.  In their terminology, therefore, a model is $\psi$-epistemic if there is even one pair of distinct quantum states that are not ontologically distinct.  A stronger notion is that of a \emph{pairwise $\psi$-epistemic} model, in which no pair of nonorthogonal pure states is ontologically distinct.

The terminology ``$\psi$-ontic'' is apt.   If preparations corresponding to two quantum states $\ket{\psi}$, $\ket{\phi}$ are always ontologically distinct, this means that the ontic state always reflects which of these states was prepared. To be physically real, it is not required that quantum states  be part of the fundamental ontology of the theory; states that supervene on the fundamental ontology are no less real for not being fundamental. An analogy: suppose that I specify some lighting and viewing conditions, and consider the set of things that, under those conditions, look yellow to me, and the set of things that, under those conditions, look blue to me.  These sets are, presumably, ontologically distinct. The two sets would not be simply describable in physical terms, and it would be difficult to explain to anyone why physical things are being lumped together in these ways without reference to the visual system of creatures like me. But they are ontologically distinct nonetheless, and the distinction reflects a distinction in reality.

On the other hand, taking ``$\psi$-epistemic''  to be simply the negation of ``$\psi$-ontic'' seems to me to be potentially misleading.  Consider, for example, a classical system, whose ontic state is represented by a point in its phase space.  Suppose that one could learn either its position, or its momentum, but not both, though it always has determinate position and momentum.  Any position is compatible with any momentum, and hence, for any position $x$ and momentum $p$, the set of ontic  states corresponding to  position $x$ overlaps with the set of states corresponding to momentum $p$. That doesn't mean that there is anything epistemic about position or momentum.

In addition, to call a model ``$\psi$-epistemic'' if there are distinct quantum states whose associated probability distributions have \emph{some} overlap, no matter how small, is potentially misleading, as it might suggest that the goal of constructing an interpretation on which quantum states are like classical probability distributions has been achieved.  This, however, would require that   the model be what has been called a \emph{maximally $\psi$-epistemic} model \citep{BCLM}.  On such a model, the indistinguishability of quantum states is fully explained by overlap of the corresponding probability distributions on ontic state space.

In addition to preparations that are perfectly distinguishable, there are also preparations that come close.  A coin-flipping procedure that yields \emph{heads} with probability very close to unity is distinguishable, not with complete certainty, but with high probability, from a coin-flipping procedure that yields \emph{tails} with probability close to unity.  For one way to quantify this, imagine the following game. A system is subjected to one of a pair of preparations, $\psi$, $\phi$, with equal probability.  You are presented with the prepared system, and are allowed to perform any experiment that you like. On the basis of the outcome of the experiment, you make a guess as to which preparation was performed.  We ask: if you choose your experiment wisely, how high can the probability of your making a correct guess be?  In the best case, there is an experiment that is certain to yield differing results depending on which preparation was applied, and the probability of correctly identifying the preparation is unity.  In the worst case, any outcome of any experiment you can do has the same probability on both preparations, and the probability of correctly identifying the preparation is no greater than one-half.  In general, the probability of correctly identifying the preparation, on an optimal strategy, is
\begin{equation}
P(\mbox{correct guess}) = \frac{1}{2}\left(1 + \sup \left|P_\psi(o) - P_\phi(o)\right| \right).
\end{equation}
Here ``sup'' means \emph{supremum}, that, is, the maximum value, taken over all outcomes $o$ of experiments that can be performed on the system in question, or if there is no maximum value but only an increasing sequence of values that approaches some limiting value, this  limiting value.  We define the \emph{distinguishability} of the preparations as
\begin{equation}
d(\psi, \phi) =  \sup \left|P_\psi(o) - P_\phi(o)\right|,
\end{equation}
The distinguishability $d(\psi, \phi)$ ranges between $0$, for the case in which $\psi$ and $\phi$ are indistinguishable, and $1$, for perfectly distinguishable preparations.    If the preparations correspond to  quantum states $\ket{\psi}$, $\ket{\phi}$, then, if there are no restrictions on the permitted experiments (that is, if every experiment that, according to quantum mechanics, is possible, is permitted), we have
\begin{equation}\label{QDist}
d(\psi, \phi) = \sqrt{1 - |\bkt{\phi}{\psi}|^2}.
\end{equation}

We will want also a notion of approximate ontological distinctness.  Given two probability distributions $P, Q$ on a measurable space $\langle \Lambda, \mc{L} \rangle$, we define the \emph{statistical distance}, also known as the \emph{total variation distance}, between $P$ and $Q$ as
\begin{equation}
\delta(P, Q) = \sup_{A \in \mc{L}} \left|P(A) - Q(A) \right|.
\end{equation}
Its value ranges between 0, when $P = Q$, and $1$, when $P$ and $Q$ have disjoint supports.  We define the \emph{classical overlap} of two probability distributions by
\begin{equation}
\omega(P, Q) = 1 - \delta(P, Q).
\end{equation}

Clearly, for any preparations $\psi, \phi$, we will always have
\begin{equation}\label{deltad}
d(\psi, \phi) \leq \delta(P_\psi, P_\phi).
\end{equation}
That is, distinguishability of two preparations can never be greater than their ontological distinctness. In the classical case, if there are no restrictions on experiments---that is, if, for any measurable subset $S$ of the state space, there is an experiment that determines whether or not the state is in $S$---then we have equality in (\ref{deltad}).  In this case, all indistinguishability of two preparations is accounted for by overlap between the corresponding state-space probability distributions.  Following \citet{BCLM}, we will say that an ontological model of some fragment of quantum mechanics is \emph{maximally $\psi$-epistemic} if and only if, for every pair of states $\ket{\psi}, \ket{\phi}$,
\begin{equation}\label{MaxEpist}
d(\psi, \phi) = \sqrt{1 - |\bkt{\phi}{\psi}|^2} = \delta(P_\psi, P_\phi),
\end{equation}
or, equivalently,
\begin{equation}
\omega(P_\psi, P_\phi) = 1 -   \sqrt{1 - |\bkt{\phi}{\psi}|^2}.
\end{equation}

\subsection{The BCLM Theorem}  Following Barrett, Calvalcanti, Lal, and Maroney (\citeyear{BCLM}) (BCLM), we define the \emph{quantum overlap} of two pure quantum states $\ket{\psi}, \ket{\phi}$ as
\begin{equation}
\omega_Q(\ket{\psi}, \ket{\phi}) = 1 - d(\psi, \phi) = 1 - \sqrt{1 - |\bkt{\phi}{\psi}|^2}.
\end{equation}
This is zero for orthogonal states ($\bkt{\phi}{\psi} = 0$), and unity when $\ket{\psi} = \ket{\phi}$.  If  one had a theory on which quantum states were like classical probability distributions, and indistinguishability of quantum states could be fully accounted for by overlap of the corresponding distributions on an ontic state space---that is, a maximally $\psi$-epistemic theory---one would have, for all preparations $\psi, \phi$ that prepare pure quantum states $\ket{\psi}, \ket{\phi}$,
\begin{equation}
\omega(P_\phi, P_\psi) = \omega_Q(\ket{\psi}, \ket{\phi}).
\end{equation}
This can be achieved for some fragments of quantum theory, which is what gives impetus to the project of attempting to construct a comprehensive theory of this sort that accounts for all quantum phenomena. However, as \citet{BCLM} demonstrate, it cannot be achieved for a model that fully reproduces quantum mechanics on a Hilbert space of dimension greater than three.

Here, in  a nutshell, is the argument. Consider a Hilbert space $\mc{H}_d$ of a dimension $d = p^n$, greater than 3, that is a power of some prime number $p$.  BCLM show that, for any $\ket{\phi} \in \mc{H}_d$, one can construct a set of state vectors, $\Psi = \{\ket{\psi_i}, i = 1, \ldots, d^2\}$   with the following properties.
\begin{enumerate}[label = \alph*)]
\item For all $\ket{\psi_i} \in \Psi$, $|\bkt{\phi}{\psi_i}| = 1/\sqrt{d}$.
\item For any pair $\ket{\psi_i}, \ket{\psi_j}$ of distinct elements of $\Psi$, either
\begin{enumerate}[label = \roman*)]
\item $\ket{\psi_i}$ and $\ket{\psi_j}$ are orthogonal to each other, and hence the corresponding preparation distributions have null overlap, or
\item the triple $\{ \ket{\phi}, \ket{\psi_i}, \ket{\psi_j} \}$ is an antidistinguishable set, and hence the corresponding preparation distributions have null joint overlap.
\end{enumerate}
\end{enumerate}
On either of the alternatives, there is no joint overlap of $\{P_\phi, P_{\psi_i}, P_{\psi_j}\}$.

Now consider the average value of the overlap  $\omega(P_\phi, P_{\psi_i})$, averaged over all elements of the set $\Psi$. Call this $\bar{\omega}(P_\phi, \Psi)$.
\begin{equation}
\bar{\omega}(P_\phi, \Psi) = \frac{1}{d^2} \sum_{i = 1}^{d^2} \omega(P_\phi, P_{\psi_i}).
\end{equation}
From the fact that no  pair  of distinct elements $P_{\psi_i}$, $P_{\psi_j}$ of $\Psi$  have non-null joint overlap with $P_\phi$, it follows that
\begin{equation}
\sum_{i = 1}^{d^2} \omega(P_\phi, P_{\psi_i}) \leq 1,
\end{equation}
and hence that
\begin{equation}\label{mean1}
\bar{\omega}(P_\phi, \Psi) \leq \frac{1}{d^2}.
\end{equation}
Since, for each  $\ket{\psi_i}$ in $\Psi$, $|\bkt{\phi}{\psi_i}| = 1/\sqrt{d}$, the value of the quantum overlap with $\phi$ is the same for each.
\begin{equation}
\omega_Q(\ket{\phi}, \ket{\psi_i}) = 1 - \sqrt{1 - 1/d}.
\end{equation}
Call this value of the quantum overlap with $\ket{\phi}$, which is the same for all members of $\Psi$, $\omega_Q(\ket{\phi}, \Psi)$.
From (\ref{mean1}), with a little bit of arithmetic, we get
\begin{equation}\label{mean2}
\bar{\omega}(P_\phi, \Psi) \leq \frac{1}{d}\left(1 + \sqrt{1 - 1/d} \right) \, \omega_Q(\ket{\phi}, \Psi) < \frac{2}{d}  \; \omega_Q(\ket{\phi}, \Psi).
\end{equation}
So, for example, for the case $d = 4$, the lowest-dimensional Hilbert space to which this theorem applies, for any vector $\ket{\phi}$ there is a set $\Psi$ of 16 vectors such that the average overlap of $P_\phi$ with distributions corresponding to elements of $\Psi$ satisfies
\begin{equation}
\bar{\omega}(P_\phi, \Psi) \leq \frac{1}{4}(1 - \sqrt{3}/2)  \; \omega_Q(\ket{\phi}, \Psi) \approx 0.47 \; \omega_Q(\ket{\phi}, \Psi).
\end{equation}
Now, the average of the overlap $\omega(P_\phi, P_{\psi_i})$ taken over the set $\Psi$ cannot be less than the smallest value of this overlap for $\ket{\psi_i}$ in that set.  Therefore, there must be at least one $\ket{\psi_i}$ in $\Psi$ such that
\begin{equation}
\omega(P_\phi, P_{\psi_i}) < \frac{2}{d} \; \omega_Q(\ket{\phi}, \ket{\psi_i}).
\end{equation}
No ontological model for quantum mechanics can come close to the dream of having quantum states be like classical probability distributions.  Even in a 4-dimensional Hilbert space, any ontological model must have, for some $\ket{\phi}$, $\ket{\psi}$, an overlap between the corresponding probability distributions that is less than half of the quantum overlap between these states. For larger Hilbert spaces, the minimum value of the ratio of classical overlap $\omega$ to quantum overlap $\omega_Q$ must be even smaller, and, for an ontological model of quantum mechanics on an infinite-dimensional Hilbert space, this ratio can have no minimum value greater than zero.  This completely dashes the hope that provides much of the impetus for the project of constructing a theory on which quantum states are not ontic.

\subsection{The PBR Theorem} The BCLM theorem applies to any ontological model of quantum mechanics, and shows that no such model can be fully $\psi$-epistemic.  There can be no theorem of this sort, which places no conditions on the ontological model, that has the conclusion that distinct quantum states are always ontologically distinct, as it is possible to construct ontological models in which any pair of distinct quantum states have some overlap in their corresponding probability distributions \citep{ABCL}.  A $\psi$-ontology theorem, therefore, must make some assumptions about the ontological model.  These assumptions should not be arbitrary, but should be physically well-motivated.  In this section we consider the  theorem of Pusey, Barrett, and Rudolph (\citeyear{PBR}) (PBR), which imposes an independence condition on probability distributions corresponding to product states.

Consider a  pair of systems, $A$, $B$, each of which is to subjected to one of two distinct quantum state preparations, $\ket{\psi}$, $\ket {\phi}$, with $|\bkt{\psi}{\phi}| \leq 1/\sqrt{2}$.  Consider now the set of states
\[
\{ \ket{\psi}_A \ket{\psi}_B, \ket{\psi}_A \ket{\phi}_B, \ket{\phi}_A \ket{\psi}_B, \ket{\phi}_A \ket{\phi}_B \}.
\]
It can be shown  that this set of states is antidistinguishable \citep{MoseleyPBR}.  That is, there is an experiment $E$ such that each outcome of the experiment is precluded by one of these preparations.  This entails that there is no four-way joint overlap between the probability distributions on ontic state space corresponding to these four states.

Now suppose we add a further postulate, the Preparation Independence Postulate (PIP).  This is actually the conjunction of two postulates.  The first postulate, which, following Leifer, we call the \emph{Cartesian Product Assumption}  (CPA), is  the condition that, when a pair of systems are independently subjected to pure-state preparations, the set of ontic states that can result from the preparation can be represented as a subset of the Cartesian product of state spaces of the individual systems.  That is, the ontic state $\lambda$ can be represented as an ordered pair $\langle \lambda_A, \lambda_B \rangle$, where $\lambda_A$ represents the ontic state of $A$ and $\lambda_B$ represents the ontic state of $B$.  The second postulate, the \emph{No Correlations Assumption},  is the condition that, for appropriate preparations,  the probability distributions corresponding to the four joint preparations are simply products of local distributions. That is, there are probability distributions $P_\psi^A$,  $P_\phi^A$ on the state space of $A$, and probability distributions $P_\psi^B$,  $P_\phi^B$ on the state space of $B$, such that, for any measurable subsets $\Delta_A$ of $A$'s state space and $\Delta_B$ of $B$'s state space, the probability, on the joint distribution $P_{\psi, \psi}$ corresponding to $\ket{\psi}_A \ket{\psi}_B$, that $\lambda_A$ is in $\Delta_A$ and $\lambda_B$ is in $\Delta_B$, is simply the product of $P_\psi^A(\Delta_A)$ and $P_\psi^B(\Delta_B)$, and the probability, on the joint distribution $P_{\psi, \phi}$ corresponding to $\ket{\psi}_A \ket{\phi}_B$, that $\lambda_A$ is in $\Delta_A$ and $\lambda_B$ is in $\Delta_B$, is  the product of $P_\psi^A(\Delta_A)$ and $P_\phi^B(\Delta_B)$, and so on, for the other possible preparations.

The NCA can itself be regarded as a conjunction of two assumptions.  The first, which we will call \emph{Ontic Parameter Independence}, is that, for a given choice of preparation on $A$, the marginal distribution of $\lambda_A$---that is, the distribution obtained from the joint distribution over $\langle \lambda_A, \lambda_B \rangle$ obtained by averaging over $\lambda_B$---is the same for each choice of preparation on $B$.  The second is the condition that, for any choice of preparations on the two systems, the corresponding probability distribution is one on which $\lambda_A$ and $\lambda_B$ are independently distributed. This assumption may well be called the No Correlations Assumption, but, since that label is already in use, we will call it the \emph{Independence Assumption}.

The Ontic Parameter Independence assumption is a causal locality assumption, and is required for compatibility with relativistic causality.  If it is violated, a choice of preparation on one system influences the probability of the result of the other preparation, even if we do not have the epistemic access to the ontic state of the system required to exploit this for signalling. The Independence Assumption is not required by causal locality, as there may be correlations between the states of the two systems that are due to influences in their common past.  The assumption really amounts to the assumption that there is some way to effect the preparations  so that such correlations are effectively screened off.  It is not required that \emph{every} procedure that we would regard as preparing the requisite quantum product state effect this screening off, only that there be \emph{some} way to do this.  Though this is not required by any sort of condition of causal locality, it is the sort of assumption that is pervasive in experimental science.

With the PIP in place, the condition that there be no four-way overlap between the four distributions considered entails ontological distinctness of $P_\psi$ and $P_\phi$   on the state spaces of the subsystems $A$ and $B$.  To see this, assume the contrary: suppose there is a subset $\Delta_A \subseteq \Lambda_A$ that is assigned nonzero probability by both $P_\psi^A$ and $P_\phi^A$, and that there is a subset $\Delta_B \subseteq \Lambda_B$ that is assigned nonzero probability by both $P_\psi^B$ and $P_\phi^B$.  Then, if the joint probability distributions satisfy the PIP, the set $\Delta_A \times \Delta _B$, which consists of pairs $\langle \lambda_A, \lambda_B \rangle$, with $\lambda_A \in \Delta_A$ and $\lambda_B \in \Delta_B$, is assigned nonzero probability by all four preparations, which is incompatible with the antidistinguishability of the set of preparations.  We therefore, conclude that either $P_\psi^A$ and $P_\phi^A$ are ontologically distinct, or $P_\psi^B$ and $P_\phi^B$ are.  However, if these are systems of the same type, subjected to the same choices of preparations, it seems reasonable to assume that the probability distributions are unchanged under an exchange of $A$ and $B$, from which it follows that  $P_\psi^A$ is ontologically distinct from $P_\phi^A$, and  $P_\psi^B$, from $P_\phi^A$.

This argument applies to any pair of states with $\ket{\psi}$, $\ket{\phi}$, with $|\bkt{\psi}{\phi}| \leq 1/\sqrt{2}$.  For a pair of distinct states with a larger quantum overlap (that is, with $|\bkt{\psi}{\phi}| > 1/\sqrt{2}$),  we  consider a larger set of systems of the same type.  Consider a system consisting of $2n$ subsystems.   Divide them into two equal subsets, which we will call $A$ and $B$. Our choice of preparations consists of a choice between subjecting all of the systems in $A$ to the $\ket{\psi}$-preparation, or subjecting all of them to the $\ket{\phi}$-preparation.  We make the same choice for $B$.
Since $\ket{\psi}$ and $\phi$ are distinct, $|\bkt{\psi}{\phi}| < 1$, and, for sufficiently large $n$, the  $n$-fold product state  $\ket{\psi}_1 \ldots \ket{\psi}_n$ has sufficiently small overlap with the $n$-fold product $\ket{\phi}_1 \ldots \ket{\phi}_n$ for the theorem to apply.  With the PIP in place, we conclude ontological distinctness of these $n$-fold product states, and, using the PIP again, of the states $\ket{\psi}_i$, $\ket{\phi}_i$ of the individual subsystems.

The result is robust under elimination of the idealization of perfect preclusion, as it must be, to be taken seriously as telling us something about the actual world.  Suppose that there exists an experiment such that, for some small $\varepsilon$,  for each outcome, there is a preparation that ascribes a probability less than $\varepsilon$ to that outcome.  On the assumption of the PIP, it follows that the overlap $\omega(P_\phi, P_\psi)$ between probability distributions corresponding to the two preparations is less than $2  \sqrt{\varepsilon}$.  See the Supplementary Information section of  \citet{PBR}  for details of the proof.

\section{Doing Without the Cartesian Product Assumption}\label{thm}
\subsection{The Preparation Uninformativeness Condition}
We have so far not discussed the status of the CPA. It is, in fact, violated in relativistic quantum field theories as we now have them.  Suppose that Alice and Bob perform operations on two systems $A$ and $B$, these operations taking place within bounded spacetime regions, at spacelike separation from each other. We assume that the effects of Alice and Bob's operations on the quantum state can be represented by operators operating on the quantum state, and  adopt the usual assumptions, required to ensure compatibility with relativistic causality,  that the operators representing Alice's operations commute  with  operators representing observables at spacelike separation from her operations, and with those representing Bob's.  On these assumptions, in the context of quantum field theory, we cannot assume that there is an operation that can be counted on to \emph{completely} remove all entanglement between the systems $A$ and $B$, and prepare them in a state that is exactly a product state.\footnote{See \citet{EntangOpenSystems} for further discussion.}  A product state can be approximated as closely as we like, but cannot be reliably achieved exactly.  In light of this, we need an independence postulate that does not presume that it is possible to prepare product states. It is best not to make any assumption about the structure of the state space at all, as we cannot expect to anticipate what sorts of state descriptions future theories might bring.

The assumption we will adopt in place of the PIP is one that we will call the Preparation Uninformative Condition (PUC). The PUC is meant to capture as much as we can of the content of the assumption that local state-preparations are possible  without presupposing anything at all about the structure of the state spaces of composite systems.  To state the assumption, we consider the following set-up.  Suppose that, for systems $A$, $B$, we have some set of possible preparations of the individual systems.  Suppose that the choice of preparation for each of the subsystems is made independently.  Following the preparation of the joint system, which consists of individual preparations on the subsystems, you are not told which preparations  have been  performed, but you are  given a specification of the ontic state of the joint system.  On the basis of this information, you form credences about which preparations were performed. In the case of ontically distinct preparations, you will be certain about what preparation has been performed; if the preparations are not ontically distinct, you may have less than total information about which preparations were performed.

We ask: under these conditions, if you are now given information about which preparation was performed on one system, is this  informative about which preparation was performed on the other? The Preparation Uninformative Condition is the assumption that it is not.   This condition is satisfied in any model that satisfies the PIP. It is also satisfied whenever the preparations are ontically distinct. In such a case, given the ontic state of the joint system, you know precisely which preparations have been performed, and being told about the preparation on one system does not add to your stock of knowledge.

One way in which the PUC can be violated is to have the ontic state space of the joint system to be the Cartesian product of the subsystem ontic spaces, and for the joint probability distributions to be ones in which the states of the subsystems are correlated.  It is also violated, as we shall see, by models, such as those constructed by \citet{ABCL}, on which nonorthogonal quantum states are never ontically distinct.

The PUC is implied by the PIP, but it is strictly weaker.  Even if the CPA is assumed, it is possible to construct models for the PBR setup, outlined in the previous section, in which the PUC is satisfied but $P_{\psi, \psi}$ and $P_{\psi, \phi}$ have nonzero overlap, which by the PBR theorem, is ruled out for models that satisfy the NCA. See \citet{MyrvoldPsiOnt} for one such construction.

The PUC, it seems to me, is a necessary condition for the operations considered to count as local state-preparations.  The substantive physical assumption made is that it is that such preparations are achievable, with  sufficient effort, or, failing that, that it is possible to achieve approximate satisfaction of the condition, to as high a degree of approximation as is desired.  This is all that is needed for the conclusions we will be drawing. It is not assumed that arbitrary operations satisfy the condition; only those that are to be counted as local state-preparations.

It should be emphasized that the PUC is \emph{not} a causal locality condition; there are operations that violate it without violating causal locality.  To see this, consider the following example, from quantum mechanics.  Consider two systems, $A$, $B$, with associated Hilbert spaces $\mc{H}_A$ and $\mc{H}_B$. Take a pair of orthogonal state vectors $\ket{0}, \ket{1}$ from each Hilbert space, and form an entangled state vector,
\begin{equation}
\ket{\Phi^+} = \frac{1}{\sqrt{2}}\left(\ket{0}_A\ket{0}_B + \ket{1}_A\ket{1}_B \right).
\end{equation}
Alice and Bob will each choose between two operations, and perform them, after which you will be told the resulting quantum state (assumed ontic for the purpose of this example). Suppose that Alice and Bob each have the choice between doing nothing, or performing a bit-flip operation that interchanges $\ket{0}$ and  $\ket{1}$.  You are told that the resulting state is just $\ket{\Phi^+}$, the same state that they started with.  You are undecided between two alternatives: either Alice and Bob both did nothing, or they both did a bit flip.  Clearly, in this situation, given the ontic state, information about  Alice's choice of operation tells you something about Bob's.  But this is a symptom of the fact that this is not  a situation that counts as a pair of local state-preparations. The systems start out in an entangled state, and remain entangled. If, on the other hand, Alice and Bob's choices are between operations guaranteed to disentangle the systems, then, given the resulting quantum state, which would be a product state, information about Alice's choice of operation would tell you nothing about Bob's.  Operations of that sort \emph{are} candidates for being regarded as local state-preparations.

\subsection{A $\psi$-Ontology result without the CPA} We consider a set-up consisting of two subsystems $A, B$, with  a choice of preparations ${\psi}$, ${\phi}$ to be made on each.  Suppose that the set of four states arising from the two choices of preparation on the two subsystems is antidistinguishable. It follows from this that there is no joint overlap between the probability distributions corresponding to the four  preparations. If, now, we impose the Preparation Uninformativeness Condition, it follows that, given the ontic state of the joint system, you will be undecided about at most one of the preparations performed on the subsystems.  That is, either the ontic state allows you to uniquely determine the preparation of $A$, or it allows you to uniquely determine the preparation of $B$.  This means that  $P_{\psi, \psi}$ and $P_{\phi, \phi}$ have null overlap, as do $P_{\psi, \phi}$ and $P_{\phi, \psi}$.

To see this, suppose the contrary.  Suppose that, given the ontic state $\lambda$, you are undecided about the preparations of both subsystems.  This indecision can obtain only when $\lambda$ is either in a joint overlap of $P_{\psi, \psi}$ and $P_{\phi, \phi}$, or in a joint overlap of $P_{\psi, \phi}$ and $P_{\phi, \psi}$.  Suppose it is the former.  Then, since there is no four-way overlap between the four preparation distributions, the ontic state must be incompatible with at least one of the other two preparations.  Suppose that it is incompatible with $\langle {\psi}, {\phi} \rangle$.  Then you are undecided between the preparations $\langle {\psi},{\psi} \rangle$ and $\langle {\phi}, {\phi} \rangle$, but have zero credence in $\langle {\psi}, {\phi} \rangle$.  Suppose, now, you are told that the $A$-preparation was ${\psi}$, and you update your credences on that information.  You have now become certain that the $B$-preparation was ${\psi}$, in violation of the PUC. Similarly, if the ontic state $\lambda$ is incompatible with $\langle {\phi}, {\psi} \rangle$, being told that the $A$-preparation was ${\phi}$ is informative about the $B$-preparation, in violation of the PUC.  The same reasoning holds,  of course,  for an overlap of   $P_{\psi, \phi}$ and $P_{\phi, \psi}$.  Therefore, from antidistinguishability of the four preparations and the PUC it follows that the ontic state $\lambda$ must uniquely determine either the quantum state of $A$ or the quantum state of $B$.

This result is robust under de-idealization.  If, for some small $\varepsilon$,  there is a 4-outcome experiment $E$ such that each preparation accords probability less than $\varepsilon$ to some outcome of $E$, then, on the assumption of the PUC,   $P_{\psi, \psi}$ and $P_{\phi, \phi}$ have small overlap, as do $P_{\psi, \phi}$ and $P_{\phi, \psi}$:
\begin{equation}
\begin{array}{l}
\omega(P_{\psi, \psi}, P_{\phi, \phi}) \leq 4 \sqrt{\varepsilon};
\\ \\
\omega(P_{\psi, \phi}, P_{\phi, \psi}) \leq 4 \sqrt{\varepsilon}.
\end{array}
\end{equation}
See \citet{MyrvoldPsiOnt} for details.

Now consider  a large number $N$ of systems, each subject to a choice of $\psi$ or $\phi$ preparations.  Call this large system, consisting of $N$ subsystems, $\Sigma_N$.  Among the experiments that can be performed on $\Sigma$ are, for each pair of subsystems $\langle i, j \rangle$, an experiment that antidistinguishes the four alternatives  $\{ \langle \psi_i, \psi_j \rangle, \langle \phi_i, \psi_j \rangle, \langle \psi_i, \phi_j \rangle, \langle \phi_i, \phi_j \rangle \}$ for those subsystems.  We require of our theory that it reproduce the quantum probabilities for any experiment that might be performed on the system.  Then, on the assumption of the PUC, this entails that the ontic state of  $\Sigma_N$ must be such that, for each pair of subsystems, this ontic state  uniquely determines the quantum state  of at least one of them.  It follows from this that the ontic state of $\Sigma_N$ must uniquely determine the quantum state of at least $N - 1$ of the subsystems. Therefore, in a large array of systems of this sort, the ontic state of the whole must uniquely determine the quantum state of the vast majority of them, with at most one exception.   By taking $N$  large enough, we can make the probability that a randomly chosen subsystem has its quantum state uniquely determined by the ontic state of $\Sigma_N$ as close to unity as we like.

Let us now impose  a \emph{Principle of Extendibility}.  This is the  requirement that the ontic state of the system $\Sigma_N$  be compatible with regarding the system as a subsystem of a larger system  $\Sigma_{N'}$ consisting of $N'$ subsystems subjected to $\psi$ or $\phi$ preparations, for arbitrarily large $N'$.  With this assumed, the probability that \emph{all} of $\Sigma_N$'s subsystems have their preparations uniquely determined by the ontic state of $\Sigma_N$ must be greater than $p$ for all $p < 1$.  That is, with probability one, all of the subsystems of  $\Sigma_N$ must be such that their quantum states are uniquely determined by the ontic state of $\Sigma_N$.

This result is, again, robust under relaxation of the assumption of perfect antidistinguishability.  It can be shown that, if, for each pair of subsystems, there is an experiment such that each of the outcomes has probability less than $\varepsilon$, then the ontic state of $\Sigma_N$ must be such that it permits almost certain identification of the quantum state of a randomly selected subsystem. Once again, see \citet{MyrvoldPsiOnt} for details.

This result holds for pure-state preparations $\ket{\psi}$, $\ket{\phi}$, with $|\bkt{\psi}{\phi}| \leq 1 /\sqrt{2}$, and, because it is robust under approximations, for any pair of preparations that approximate such states closely enough to permit approximate antidistinguishability.  For quantum states with a greater quantum overlap, we can use the result to show that   for sufficiently large $n$, the $n$-fold product $\ket{\psi}_1 \ldots \ket{\psi}_n$ is ontologically distinct from $\ket{\phi}_1 \ldots \ket{\phi}_n$.   Unlike the PIP, the PUC does not permit us to conclude, straightaway, that for individual subsystems $\ket{\psi}$ is ontologically distinct from $\ket{\phi}$.  But it is hard to believe that there is a theory worth taking seriously which is such that  $\ket{\psi}$ and $\ket{\phi}$ are ontologically distinct whenever $|\bkt{\psi}{\phi}| \leq 1 /\sqrt{2}$ and, though    $\ket{\psi}$ and $\ket{\phi}$ are not ontologically distinct for individual systems, for a sufficiently large collection of systems the set of states that can arise from subjecting all of them  to the $\ket{\psi}$-preparation is ontologically distinct from the set of states that care arise from subjecting all of them to the $\ket{\phi}$-preparation.  To echo \citet{FVLC}, if someone presents me with a candidate for such a theory, I will not refuse to listen, but I will  not myself try to make such a theory.

\section{Conclusion} The PUC is a fairly weak condition,  consistent with pervasive nonseparability of state descriptions, and satisfied even in manifestly nonlocal theories of quantum phenomena, such as the de Broglie-Bohm theory.  Though, of course, it is  possible to consider theories on which it does not hold, it must be admitted that we have \emph{no evidence whatsoever} that this condition is not satisfied in the actual world.  Any theory that satisfies this condition must have it that distinct states with an inner product not greater than  $1 /\sqrt{2}$ are ontologically distinct. Furthermore, no theory whatsoever, whether it satisfies the PUC or not, can both reproduce the quantum probabilities for results of experiments and satisfy the \emph{desideratum} that indistinguishability of states  be fully accounted for by overlap of the corresponding probability distributions.  For these reasons, though the project of constructing a theory, of the sort envisaged by Einstein,  in which quantum states are analogous to  probability distributions in classical statistical mechanics, was well-motivated, we have to admit that the fruit it has borne consists of  insight into why the goal cannot be achieved.

We have reached the point, it seems to me, at which anyone concerned with understanding what the empirical success of quantum theory is telling us about the world should acknowledge that it is telling us that the furniture of the world includes something  corresponding to quantum states.  This, of course, does not come close to settling the question of what a complete account of the world might be like, and the old questions remain about how best to understand what it is that quantum theory is telling us about the world we live in.

\subsection*{Acknowledgments}
Support for this research was provided by Graham and Gale Wright, who sponsor the Graham and Gale Wright Distinguished Scholar Award at the University of Western Ontario. This research was also supported by Perimeter Institute for Theoretical Physics. Research at Perimeter Institute is supported by the Government of Canada through the Department of Innovation, Science and Economic Development and by the Province of Ontario through the Ministry of Research and Innovation. I am grateful to the members of the Philosophy of Physics Reading Group at the University of Western Ontario for helpful feedback and discussions.

\newpage
\bibliographystyle{chicago}

\end{document}